# Crossover Scaling of Wavelength Selection in Directional Solidification of Binary Alloys

Michael Greenwood, Mikko Haataja, and Nikolas Provatas

*Department of Materials Science and Engineering, and Brockhouse Institute for Materials Research,*
*McMaster University, 1280 Main Street West, Hamilton, Ontario, Canada, L8S 4L7*

(Dated: March 20, 2004)

We simulate dendritic growth in directional solidification in dilute binary alloys using a phase-field model solved with an adaptive-mesh refinement. The spacing of primary branches is examined for a range of thermal gradients and alloy compositions and is found to undergo a maximum as a function of pulling velocity, in agreement with experimental observations. We demonstrate that wavelength selection is unambiguously described by a non-trivial crossover scaling function from the emergence of cellular growth to the onset of dendritic fingers, a result validated using published experimental data.

PACS numbers:

Pattern formation in solidification is a paradigm fundamental to many problems of scientific and industrial relevance. In commercially cast alloys the process of solidification sets in the fundamental length scales that characterize their microstructure, which in turn is largely responsible for their mechanical properties [3, 8]. This is particularly true in emerging technologies such as strip-casting in which alloys are rapidly cooled directly into thin strips and undergo little thermomechanical processing after being cast. For such processes the attainment of desired microstructure must be achieved largely through the physics of solidification.

Dendritic structures form the basic units of solidification. At low undercooling the growth rate of isolated dendrites is selected by a solvability criterion established by a singular perturbation in the surface tension anisotropy [5, 19], a result confirmed by asymptotically matched phase-field simulations [18, 30]. At high undercooling, however, recent phase-field simulations using data from atomistic simulations [13, 14] suggest that dendritic morphology and growth rates are strongly influenced by interface kinetics [6].

In most casting applications solidification occurs as competitive growth of multiple dendritic arrays growing as an advancing front, directionally solidified through a thermal gradient established by the cooling rate. A useful paradigm used in the study competitive dendritic growth –particularly in thin-strip casting– is 2D directional solidification. In this process a binary alloy in a thin film geometry is solidified through a fixed temperature gradient $G$, with the cooling rate controlled by a pulling speed $V$. After becoming unstable via the Mullins-Sekerka instability [28] the solidification front develops a variety of complex cellular and dendrite patterns. The scale and morphology of these dendritic patterns controls the microstructure (and thus its properties) and solute segregation of the solidified alloy.

The behaviour of competitive growth of dendritic or cellular arrays has been well-characterised experimentally using alloys of organic analogues of metals such as succinonitrile (SCN) or pivalic acid (PVA) [2, 4, 11, 21, 24–27, 34–36] and theoretically [2, 7, 9, 12, 15, 20, 22–24, 29, 32, 33]. Several theories have been proposed for predicting this length scale selection in dendritic and cellular arrays. These often use simple geometrical arguments that relate the tip shape to the fastest linearly unstable wavelength, which in turn is related to process parameters (thermal gradient and velocity) [7, 15, 23, 24, 35]. While some predictions are in qualitative agreement with experimental trends [7, 23], they display a marked quantitative discrepancy from experiments. In particular, theoretical predictions are typically validated by fitting the data over certain ranges of pulling velocity [24, 35]. However, such procedures can be seriously hampered by the limited range of data and/or crossover effects [24].

Crossover phenomena can typically be attributed to a competition between two or more physical mechanisms operating across different scales (time, length, velocity etc.). A very powerful approach that has been successfully employed in capturing crossover phenomena in diverse physical systems is that of scaling. Here, one first attempts to isolate the material/process-dependent scales, and the generic (i.e., universal) behavior of the particular system emerges as a scaling collapse of the data once it has been properly non-dimensoinalized by these scales. The attractive feature of the scaling approach is that, when successful, it quantitatively describes the behavior of the system over many scales.

In this Letter, we report simulations that examine length scale selection in directional solidification. We use a phase-field model solved on an adaptive grid [30, 31], gaining access to system sizes several orders larger than the diffusion length and to greatly reduced simulation times, which helps in obtaining convergence toward steady state conditions. We show that wavelength versus velocity is unambiguously described by a crossover scaling function spanning the *entire* range from the emergence of cellular growth to the onset of dendritic fingers. Scaling collapse is obtained by defining a new dimensionless velocity and wavelength based on physical length scales of the system. Furthermore, this novel scaling function also describes quantitatively the scaling collapse of previously published experimental data from Ref [24].

Directional solidification was simulated using a recent phase-field formulation for binary alloy developed in Ref. [17]. The model describes the solidification of a dilute binary alloy with liquid concentration ratio defined by a partition coefficient $k$. The model couples an order parameter $\phi$ to a concentration field $C$. The field $\phi(\vec{x})$ takes on the values $\phi = -1$ in the solid phase, $\phi = 1$ in the liquid phase and interpolates continuously between these states in the interface region. In units where space is scaled by $W_o$, the interface width, and time by $\tau_o$, the interface kinetics time, the equations of motion for the two fields are given by

$$\frac{\partial C}{\partial t} = -\nabla \cdot \vec{j}$$

and,

$$\begin{aligned} A^2(\vec{n})\frac{\partial \phi}{\partial t} &= \vec{\nabla} \cdot (A^2(\vec{n})\vec{\nabla}\phi) + \phi(1-\phi^2) \\ &\quad - \frac{\lambda}{1-k}(e^\mu + \theta - 1)(1-\phi^2)^2 \\ &\quad + \frac{1}{2}\vec{\nabla} \cdot \left[|\nabla\phi|^2 \frac{\partial A^2(\vec{n})}{\partial(\vec{\nabla}\phi)}\right] \end{aligned}$$

where $e^\mu = 2(C/C_l^o)/(1+k-(1-k)\phi)$ and the flux $\vec{j} = -DCq(\phi)\nabla u - a_t C_l^o(1-k)e^\mu(\partial_t\phi)\vec{n}$ with $\vec{n} = \nabla\phi/|\nabla\phi|$, the unit normal to the contours of $\phi$. The pulling velocity $V_p = V_s\tau/W$, where $V_s$ is the dimensional pulling speed, and $C_o^l$ is the liquid phase alloy concentration. Dimensionless temperature is defined by a *frozen field* $\theta = (1-k)(z - V_p t)/l_T$, where $z$ is the pulling direction, $l_T = |M_L|(1-k)C_o^l W_o/(G\lambda)$ is the thermal length and $G$ the dimensional thermal gradient. The constant $M_L$ is the liquidus slope. The concentration and phase fields are coupled via the constant $\lambda$. The dimensionless diffusion constant is $D = D_L \tau_o/W_o^2$ where $D_L$ is the diffusion constant in the liquid which sets the diffusion length $l_D = 2D/V_p$. Two-sided diffusion is controlled by the function $q(\phi) = (1-\phi)/(1+k-(1-k)\phi+(1+\phi)\xi/2$ where $\xi = D_s/D = 10^{-4}$. Surface tension anisotropy is defined in terms of $\vec{n}$. Specifically, $A(\vec{n}) = [1-3\epsilon_4][1 + \frac{4\epsilon_4}{1-3\epsilon_4}((n_x)^4 + (n_y)^4)]$, where $\epsilon_4$ is the anisotropy constant. The anisotropic interface width is thus defined as $W(\vec{n}) = W_o A(\vec{n})$ and the characteristic time $\tau(\vec{n}) = \tau_o A^2(\vec{n})$ [17, 18, 30].

The constants $W_o$, $\tau_o$, $\lambda$ and $a_t$ are inter-related by an asymptotic analysis [17] which maps the phase-field model onto the sharp interface limit defined by: (1) solute diffusion in the bulk phases, (2) flux conservation at phase-boundaries and (3) the Gibbs-Thomson condition $C_{int} - C_{eq} = -d(\vec{n})\kappa - \beta(\vec{n})V$, with $\kappa$ the local interface curvature, $d(\vec{n}) = d_o[A(\vec{n}) + \partial^2 A/\partial(\cos^{-1} n_x)^2]$ where $d_o$ is the isotropic capillary length, and $V$ is the local interface speed. In the limit $\beta = 0$, we obtain $d_o/W \approx 0.8839/\lambda$, $D \approx 0.6267\lambda$ and $a_t = 1/(2\sqrt{2})$. This phase-field formulation has the distinct advantage of quantitatively equivalent to the sharp-interface model and still allowing a flexible choice of diffusion functions $q(\phi)$ and a large choice of interface width $W$. Other thermodynamically consistent formulations are also possible [10], although they admit somewhat more stringent choices of diffusivity functions.

The phase-field model was simulated using a finite element method on an adaptive grid, with zero-flux boundary conditions in both $C$ and $\phi$ as in Ref. [31]. Solidification is initiated by a small-amplitude, randomly perturbed solid/liquid interface. The initial solute profile $C(\vec{x}, 0)$ was set to a steady-state diffusion profile normal to the interface, while $\phi(\vec{x}, 0) = \tanh(\vec{x}/\sqrt{(2)})$ along the normal to the interface.

All system sizes were $6400$ in the $z$-direction and varied from $2048 - 6400$ in the transverse direction. The minimum grid spacing was set to $dx_{min} = 0.39$ in all cases. We used explicit time integration, with a time step $dt = 0.008$ as in Ref. [17]. The anisotropy $\epsilon_4 = 0.002$. The coupling parameter varied from $\lambda = 1.3$ for high velocities to $\lambda = 20$ for low velocities near the cellular onset.

We simulated directional solidification for a PVA-0.1%ACE system with a partition coefficient $k = 0.16$. Three sets of parameters $(G, \lambda, D, d_0/W)$ were considered: $(10K/mm, 20, 12.5, 0.044)$, $(660K/mm, 3, 1.88, 0.295)$, and $(1500K/mm, 1.3, 0.8147, 0.68)$. We note that using $\lambda = 20$ enabled very long-time runs close to the planar-to-cellular onset boundary. As the pulling velocity $V_p$ (oriented along the $z$-axis, parallel to the surface tension anisotropy) was varied we observed cellular structures at low values of $V_p$, while at high velocities we observed the emergence of dendritic morphologies, as shown in Figure 1. A measure of the primary branch spacing in the computed data was obtained by examining the power spectrum of the solid-liquid interface profile as a function of time. The main peak position, which corresponds to the visually observed primary spacing $\lambda_1$, was computed using the definition $k_{mean} = \frac{\sum_{i=0}^n k_i P_i}{\sum_{i=0}^n P_i} \equiv 2\pi/\lambda_1$. The value $k_{mean}$ was plotted versus inverse time and extracted to 0 to obtain an estimate of $\lambda_1$. Data for $k_{mean}$ vs. $1/t$ is shown in Figure 1 for four different velocities corresponding to the $\lambda = 20$ data. Most importantly, we note that the main contribution to $\lambda_1$ settles in reasonably fast, while the tip radius evolves on a much longer time scale.

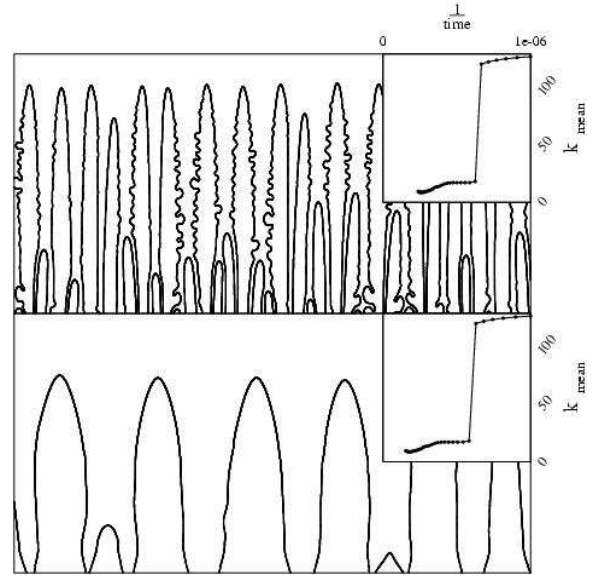

FIG. 1: *(bot) Typical cellular structures for $V = 150\mu m/s$ and $G = 1500K/mm$. (top) Dendritic morphology with sidebranch structures corresponding to $V = 90\mu m/s$ and $G = 10K/mm$. The insets show the peak wavelength versus inverse time which is used to extract the long-time branch spacing.*

Figure 2 shows the primary branch spacing $\lambda_1$ for our com-puted data. For two of our data sets a maximum occurs in $\lambda_1$ as $V_p$ approaches the planar-cellular onset. Furthermore, we found that this maximum value occurs at $V^*$ satisfying $l_T \approx l_D$. The presence of such a maximum has been been predicted theoretically [23] and observed in several experiments [4, 21, 24]. As a comparison, the data from Ref. [24] is shown in the inset of Fig. 2. The three experiments shown are for SCN-0.25mol%Salol at 13K/mm, SCN-0.13mol%ACE at G=13K/mm and PVA-0.13mol%Ethanol at G=18.5K/mm. We note that certain experiments in similar systems do not show the peak in $\lambda_1$ [35]. The reasons for this are not clear although plausible explanations are 3D effects due to the thickness of the sample slides or a long-lived metastable state at velocities near the onset [1]. For $V_p > V^*$, the data in Fig. 2 displays the characteristic monotonically decreasing wavelength as a function of velocity.

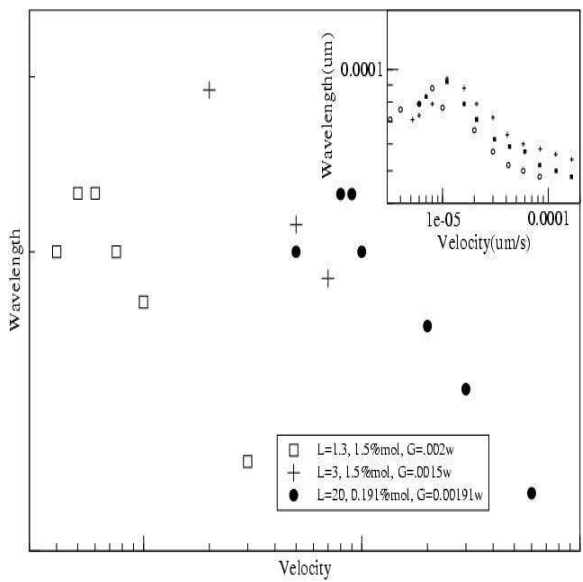

FIG. 2: *Computed dendrite spacing from simulated parameters listed in the text. The inset shows experimental dendrite spacing data obtained from digitized images from Ref. [24]. The data is for alloys of Succinonitrile and Pivalic Acid Crystals grown in a unidirectional thermal gradient.*

There has been a great deal of work on scaling relationships for primary (and other) branch scales in different morphological regimes (See [34] and references therein). These typically take the form $\lambda_1 = A l_T^\alpha l_D^\beta d_o^\gamma$, where $A$ is constant independent of the physical length scales. The prefactor and exponents $\alpha$, $\beta$ and $\gamma$ can vary depending on the semi-empirical and/or geometrical arguments of a given theory [15, 21, 23, 34]. Moreover, the scaling form must necessarily assume distinct exponents when different growth regimes are present [21].

We describe the primary scaling selection by proposing a crossover scaling function of the form

$$\frac{\lambda_1}{\lambda_c} = \frac{l_D}{l_T} f\left(\frac{l_T}{l_D}\right)$$

where $\lambda_c$ is a characteristic length scale that depends on the $l_T$, $l_D$ and $l_{TR}(V_c)$, where $V_c$ is the velocity at the planar-to-cellular onset and $l_{TR}(V_p)$ is the velocity-dependent thermal length determined by the intersection of the temperature field and the steady state solute field that that follows the liquidus line exactly. In the limit $V_p \rightarrow \infty$, $l_{TR}(V_p) \rightarrow l_T$ becomes the normal constitutional supercooling criterion. The length scale $l_{TR}(V_p)$ essentially determines the extent to which the cellular fingers can grow [16]. Figure 3 shows our computed data collapsed onto a scaling function of the form above. Also shown on the scaled plot is the experimental data from Refs. [21, 24], which also exhibits a maximum in $\lambda_1$. In each case $\lambda_c$ was selected so as to obtain the best data collapse. Data collapse of the experiments and simulations is obtained by plotting $(\lambda_1 l_D)/(\lambda_c l_T)$ against $(l_T/l_D - l_T^c/l_D^c)$, where the $l_T^c \equiv l_{TR}(V_c)$ and $l_D^c$ are the thermal and diffusion lengths evaluated at the cellular to planar onset, respectively. The plot is remarkable in that it predicts a scaling function describing the wavelength versus velocity over a range of pulling velocities, thermal gradients and alloy concentrations. For the data in this work the crossover function of Fig. 3 covers the regime from cellular regime and crosses over into the dendritic regime. Extension of the crossover scaling function further into the dendritic regime are currently being investigated. The linear fits correspond to best fit of the data showing two distinct power-law regions for the two branches of $\lambda_1$, and a crossover region between the regimes.

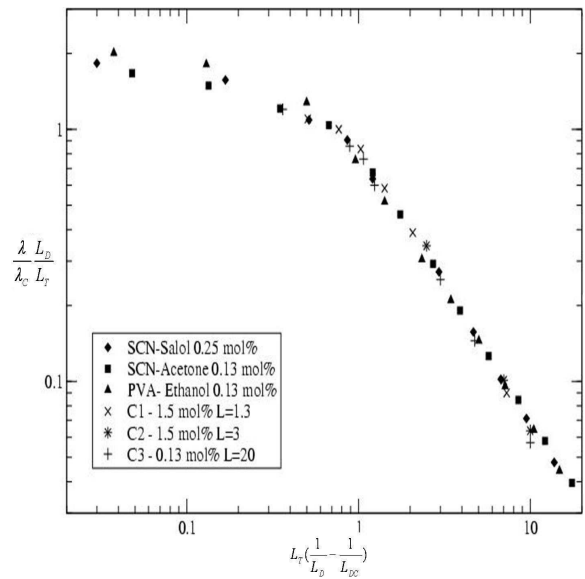

FIG. 3: *Experiments [21, 24] and computed data for SCN and PVA scaled to material properties, producing a single scaling function. See text for details.*

Figure 4 shows a plot $\lambda_c$ vs. $R/d_o$, where $R=\sqrt{\lambda_{ms} l_{TR}(V_c)}$ is proportional to the wavelength at the planar-to-cellular onset predicted by Kurz and Fisher [23]. The data suggests a form of $\lambda_c$ given by $\lambda_c = d_o(1 + \sqrt{\lambda_{ms} l_{TR}(V_c)}/d_o)$. On the same figure is also plotted $\lambda_c$ Vs. $(d_o l_T l_D)^{1/3}$ [34]. Over the range of the experiments and our computations Fig. 4 shows that for large values of $\lambda_c/d_o$, the characteristic scale $\lambda_c$ is consistent with both theoretically predicted length scales. At smaller wavelength the characteristic length at the onset displays non-trivial corrections from theoretical predictions.

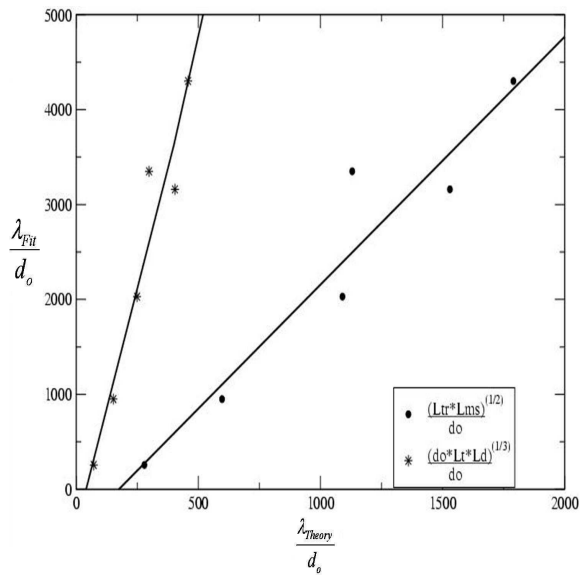

FIG. 4: $\frac{\lambda_{fit}}{d_o}$ vs $\frac{\lambda_C}{d_o}$. A plot of $\lambda_c$ Versus two theoretically predicted forms for $\lambda_1$. The first, $\sqrt{\lambda_{ms} l_{TR(V_c)}}$ (lower line), is determined by a geometric relationship assuming an elliptical tip, and the second (steeper line) through the geometric mean of the three lengths scales in directional solidification [34] $(d_o L_D L_T)^{\frac{1}{3}}$; both scales are in units of $d_o$

To summarize, we have simulated wavelength selection of cellular patterns in 2D directional solidification using the phase-field method solved on an adaptive grid. The selected wavelength displays non-monotonic behavior as a function of pulling speed; in particular, it displays a maximum for intermediate values of $V_p$. An intriguing feature of the computed data is that they can be collapsed onto a single crossover scaling function. Furthermore, it was shown that this novel scaling function also describes wavelength selection in previously published experimental data from Ref. [24]. We strongly believe that the scaling approach undertaken in the present work can be further developed into a predictive tool for microstructure selection in solidification processing.

*Acknowledgments*– This work was supported by the Center for Automotive Materials and Manufacturing at McMaster University. We thank SHARC-NET and the Materials and Manufacturing Research Institute of McMaster University for supercomputer time.